\documentclass[conference]{IEEEtran}
\usepackage{etex}
\IEEEoverridecommandlockouts
\usepackage{cite}
\usepackage{amsmath,amssymb,amsfonts}
\usepackage{algorithmic}
\usepackage{graphicx}
\usepackage{textcomp}
\usepackage{oz}
\usepackage{subfig}
\usepackage{amssymb}
\def\BibTeX{{\rm B\kern-.05em{\sc i\kern-.025em b}\kern-.08em
    T\kern-.1667em\lower.7ex\hbox{E}\kern-.125emX}}
\begin{document}

\title{Understanding and Formalizing Accountability\\ for Cyber-Physical Systems
\thanks{This work was supported by the  Deutsche Forschungsgemeinschaft (DFG)
under grant no. PR1266/3-1, Design Paradigms for Societal-Scale Cyber-Physical
Systems. This article was accepted in the IEEE Conference on Systems, Men and
Cybernetics and is \copyright 2018 IEEE.}
}

\author{\IEEEauthorblockN{Severin Kacianka}
\IEEEauthorblockA{\textit{Technical University of Munich} \\
\textit{Chair for Software and Systems Engineering}\\
Munich, Germany \\
severin.kacianka@tum.de}
\and
\IEEEauthorblockN{Alexander Pretschner}
\IEEEauthorblockA{\textit{Technical University of Munich} \\
\textit{Chair for Software and Systems Engineering}\\
Munich, Germany \\
alexander.pretschner@tum.de}
}

\maketitle

\begin{abstract}
Accountability is the property of a system that enables the uncovering of causes
for events and helps understand who or what is responsible  for these events. 
Definitions and interpretations of accountability differ; however, they are
typically expressed in natural language that obscures design decisions and the
impact on the overall system. 
This paper presents a formal model to express the accountability properties of
cyber-physical systems.  To illustrate the usefulness of our approach, we
demonstrate how three different interpretations of accountability can be
expressed using the proposed model and describe the implementation implications
through a case study. 
This formal model can be used to  highlight context specific-elements of
accountability mechanisms,  define their capabilities, and  express different
notions of accountability. In addition, it makes design decisions explicit and
facilitates discussion, analysis and comparison of different approaches.
\end{abstract}

\begin{IEEEkeywords}
CPS, STS, accountability, formal model
\end{IEEEkeywords}

\section{Introduction}

Cyber-physical systems (CPS), such as robots, drones, vehicles, and industrial
control systems, use sensors, software, and actuators to sense, monitor, and
control or influence the physical world. CPSs cannot be tested completely and,
differing from information systems, mistakes made by such systems are always
permanent.  It is impossible to ``undo'' actions of CPS or roll the world back
to the last known good state.  The primary problem is that such systems are
\emph{open}, i.e., they work jointly with other systems that are not known
a~priori. This lack of clearly defined system boundaries results in unexpected
interactions and problems. Thus it is important that such systems are
\emph{accountable}: They should provide evidence of their actions that an
investigator can understand,  and the gained knowledge  can then be used
to improve the system, trace bugs, or as forensic evidence in legal
actions~\cite{datta}. An \emph{accountability mechanism} will not prevent
unwanted events; however, it will help to detect them and determine their root
cause. In addition, the analysis can be applied to improve the system and
develop preventive measures.  For example, consider a recent fatal accident
involving a self-driving Uber car \cite{ubercrash} (Section~\ref{sec:example}).
For a currently unknown reason the autonomous car crashed into and killed a
pedestrian who was crossing the street at night.  In this example,
accountability does not automatically imply  legal liability or even punishment
for Uber or the safety driver. Here, accountability is the requirement to
explain the chain of events that led to this tragic death. The results may then
be used to infer legal liability and  improve the system to avoid such accidents
in the future.   However, accountability is a context specific concept and will
differ in various contexts. For example, an autonomous car will be held to a
different standard than a toy helicopter.  Many different notions of
accountability exist in various domains, and no unified framework exists to
capture their differences and evaluate their impact on the implementation of
CPSs. 

In this paper, we investigate the \emph{problem} of capturing the relevant
accountability properties of a CPS. As a \emph{solution}, we propose a formal
model of accountability that allows us to express and compare notions of
accountability.

Our primary \emph{contribution} is the formalization of three notions of
accountability. In addition, we demonstrate how these formalizations can be used
to compare such notions and analyze their impact on the implementation of a CPS
in a case study inspired by the self-driving Uber car incident.

\section{Background}

\subsection{Accountability in Computer Science}

In computer science, accountability came into focus following a study published
by Weitzner et al. \cite{Weitzner:2008}, who looked at preventing data leaks in
information systems (e.g.,  medical record systems) and proposed abandoning the
classic preventive approach to data privacy in favor of a detective approach.
Rather than attempting to  prevent data leaks, they suggested systems be built
such that leaks can be easily identified.  Their approach relies on existing
social measures (e.g., courts) to punish perpetrators and deter misbehavior.
Feigenbaum et al.~\cite{feigenbaum2011towards} and K\"{u}sters 
et al.~\cite{kusters2010accountability} provided detailed formal definitions of
specific interpretations of accountability in particular contexts. The problem
with such approaches is that they adopt a single interpretation of
accountability and build their system around it. However, accountability  is
highly dependent on  social and cultural contexts. Thus, any system that claims
to be accountable must necessarily support different notions of accountability
depending on the context. To validate this assumption, we recently conducted a
systematic mapping study \cite{mapping} and found that implementations of
accountability mechanisms in the literature are very diverse and do not follow a
unified model or interpretation of accountability. 

The notion of determining causes and attributing responsibility is not a new
idea and is not always referred to as accountability. For example, programs are
instrumented with logging statements \cite{amir2016correct} to audit and
reconstruct their execution. In computer security, developing systems to gather
relevant evidence during  cyber attacks is referred to as forensics-by-design
\cite{fbd2016}. Runtime verification \cite{leucker2009brief} is a collection of
techniques to check whether a system violates some correctness property. As with
current approaches to accountability, such runtime verification techniques  do
not lend themselves to the expression and comparison of  system's accountability
relative to its socio-technical contexts.

\subsection{Accountability in Social Sciences}
\label{sec:social}

As an example for a definition of accountability, 
Lindberg~\cite{lindberg2013mapping} surveyed the social science literature  and
provided the following synthesized definition of accountability:
\begin{enumerate}
\item An agent or institution who is to give an account (A for agent);
\item An area, responsibilities, or domain subject to accountability (D for
domain);
\item An agent or institution to whom A is to give account (P for principal);
\item The right of P to require A to inform and explain/justify decisions with 
regard to D; and 
\item The right of P to sanction A if A fails to inform and/or explain/justify 
decisions with regard to D.
\end{enumerate}

However, the seeming clarity of the definition hides many small differences and
one major point: The definitions do not agree on whether  a principal should
have the right to sanction agents for the content of their account, or only if
they fail to provide a justification for their decisions.  

In psychology, Hall et al. \cite{hall2017psy_acc} surveyed \emph{felt
accountability} literature and found that accountability is generally understood
to mean that actors think there is a possibility that their actions will be
evaluated by a third party. 

\subsection{Accountability in Organizations}
\label{sec:raci}
In contrast to social sciences,  organizational sciences often apply a
Responsible-Accountable-Consult-Inform (RACI) framework \cite{smith2005role} to
visualize the roles of people in an organization. The elements of the framework
are described as follows:
\begin{itemize}
\item \textbf{R}esponsible: The individual who completes a task. Responsibility
can be shared.
\item \textbf{A}ccountable: The person who answers for an action or decision.
There can be only one such person.
\item \textbf{C}onsult: Persons who are consulted prior to a decision. 
Communication must be bidirectional. 
\item \textbf{I}nform: Persons who are informed after a decision or action is
taken. This is unidirectional communication.  

\end{itemize}

Note that the RACI framework focuses on the organization rather than individuals
in the organization. While differing from Lindberg's definition 
\cite{lindberg2013mapping}, this perspective can be very useful to model the
accountability of CPSs that work in conjunction with humans such as service
robots.  We modify the RACI framework as follows to make it more appropriate for
CPS contexts.

\begin{itemize}
\item \textbf{R}esponsible: The entity whose action(s) caused an outcome.
\item \textbf{A}ccountable: The person/institution who knows about the machine's
actions and has the ability to change its action.
\item \textbf{C}onsult: Persons/institutions who build and set-up the
machine(s).
\item \textbf{I}nform: Persons/institutions who know about the system and are 
tasked with after-the-fact analysis of an outcome.    

\end{itemize}

Relative to this modified definition, many studies, notably Feigenbaum et al.
\cite{feigenbaum2011towards}, have defined accountability more in line with what
is the \emph{responsible} element of the RACI model. While it is important to
know which machine caused a given outcome, stopping at responsibility is often
insufficient, because we cannot reasonably punish a machine. We want to make it
easy to identify which person or institution can be held accountable for the
action of the machine.

\section{Formal Model} 
\label{sec:model}

\subsection{Overview}

To help express and discuss accountability concepts, we developed the following
formal model, in which we employ Z notation \cite{spivey}. Initially, the
proposed model introduces some \emph{given sets}. These sets represent the basic
concepts of our universe of discourse and they are \emph{underspecified}. In
other words, we leave it to the concrete implementation to define a suitable
representation for them. Next, the proposed model describes some
context-specific \emph{interfaces} that are implementation-specific and thus are
also underspecified.  Based on these interfaces, we describe some basic
\emph{axioms} that are valid independent of the concrete system. We conclude
with a description of a \emph{state space} suitable for CPSs and provide
detailed examples in Section~\ref{sec:example}.

\subsection{Given Sets}

\begin{footnotesize}
\begin{zed}
   [ Account, Action, Component ] \\
   Being ::= Human | Animal\\
   Event ::= EnviromentEvent | SystemEvent\\
   Principal ::=  Person | LegalEntity \\
\end{zed}
\end{footnotesize}
Here, an \texttt{Account} can, for example, be a well-structured log file or a
story told by a human witness, and \texttt{Actions} are anything a principal can
do to prevent a CPS from doing something. How \texttt{Components} look, are
scoped, and implemented depends on the given use case. \texttt{Beings} are
entities that the system may encounter; however, \texttt{Beings}  have no
control over the system and do not play an active part. \texttt{Events} can
either be \texttt{SystemEvents} or \texttt{EnviromentEvents}. The sources of
\texttt{EnviromentEvents} are external to the system, and such events will not
show up in log files directly. In contrast, \texttt{SystemEvents} are recorded
in log files.  Furthermore, \texttt{Principals} can either be \texttt{Persons}
or \texttt{Legal\-Entities}, e.g., companies. 

The distinction between \texttt{Principal} and \texttt{Being} is not obvious.
While persons are generally humans, in this context, we differentiate them in
terms of their function, where a \texttt{Principal} is an entity that is part of the
system, its legal ``surroundings'' and can be responsible for actions of the
system, and a \texttt{Being} is not an active component of the system. Note that
humans can be both principals and beings.

\subsection{Context- and Implementation-specific Interfaces}
Next, we define a few axiomatic relations. Similar to the basic types above, we
leave the details to the specific implementation. We simply require them to
exist and comply with the given signature.

\begin{footnotesize}
\begin{axdef}
Observation: (Event \cross Component) \fun  Account\\
ComponentConfiguration: Component \fun Principal\\ 
hasAccount:  Account \rel  Principal\\
correctionAction:   (Principal \cross Component) \fun  Action\\ 
caused:  Event \pfun  \power(Component) \\
 \end{axdef}
\end{footnotesize}
An \texttt{Observation} indicates that an event caused by or about a component
was observed by someone or something and is part of their account.
\texttt{ComponentConfigurations} track if a component was set up and/or
configured by some principal. \texttt{hasAccount} is a relation that should
contain all accounts belonging to a principal. Note that we cannot assume to
know all such accounts  because a principal may lie or forget.
\texttt{correctionAction} gives the set of all actions a principal can take to
correct the behavior of a component. \texttt{caused} returns the set of
components directly responsible for an event. In some cases, this can be
computed efficiently \cite{accbench}. The cause of an event may be unknown; thus
it is a partial function. Furthermore, this function could point to multiple
possible causes. Therefore we require a dedicated resolution process to resolve
such disagreements.

\subsection{Context-independent Axioms}

\begin{footnotesize}
\begin{axdef}
   informed:  Component  \fun   \power(Principal)
\where
  	\forall c: Component @ 
  		\forall p : Principal @ \\
  		p \in informed(c)  \iff \\ 		
  		 \exists e : Event  @   hasAccount(Observation(e,c)) = p \\ 
\end{axdef}
\begin{axdef}
   constructed:  Component  \fun   \power(Principal)
\where
  	\forall c: Component @ 
  		\forall p : Principal @ \\
  		p \in constructed(c)  \iff \\
  			\exists c : Component @ ComponentConfiguration(c) = p
\end{axdef}
\begin{axdef}
   responsible:   Component  \rel  Principal
\where
  	\forall c: Component @ 
  		\forall p : Principal @ \\ 	
  			 	responsible(c) = p \iff \\
  			 	 p \in informed(c) \land (p,c) \in dom(correctionAction)		 	
\end{axdef}
\end{footnotesize}
\texttt{informed} provides the set of all principals that gain knowledge
about a component. This means that the principal has an account in which an
observation about a component that is somehow responsible for this event is
logged. A principal has helped \texttt{construct} a system if they worked on
the  \texttt{Component\-Configuration}. The notion of \texttt{responsible} is
closely linked to the notion of \texttt{caused} above. However, while causality
is a much broader concept, \texttt{responsible} means actual people who can stop
someone or something from doing something. A principal is \texttt{responsible}
for some component, if it knows about it and could do something about it.

\subsection{Definitions of Accountability}

With these basic building blocks, we define three notions of accountability:

\begin{footnotesize}
\begin{axdef}
   raci\_accountable:   Event  \rel  Principal
\where
  	\forall e: Event @ 
  		\forall p : Principal @ \\ 	
  			raci\_accountable(e) =  p \iff \\
  				\exists c : Component @ c \in caused(e) \land responsible(c) = p 	
\end{axdef}
\begin{axdef}
   lindberg\_accountable:   Component  \rel  Principal
\where
  	\forall c: Component @ 
  		\forall p : Principal @ \\ 	
  			lindberg\_accountable(c) = p \iff \\
  					 p \in informed(c) \land\\ 
					 (responsible(c) = p \lor p \in constructed(c))
\end{axdef}
\begin{zed}
 hall\_accountable == \{c: Component | \\
 \exists  p : Principal @ p \in informed(c)  @ c \}
\end{zed}

\end{footnotesize}

First, a principal is \texttt{raci\_accountable} (Section~\ref{sec:raci}) for
some event if a component for which the principal is responsible caused the
event.  The second definition expresses Lindberg's definition of accountability
\cite{lindberg2013mapping} (Section~\ref{sec:social}). A component is
\texttt{lindberg\_accountable} to a principal if: 
\begin{itemize}
\item The component must give an account to a principal. In our model this means
that the principal is \texttt{informed} about the component.
\item There is an area of accountability. The area of accountability is
implicitly defined by the purpose of the CPS. 
\item A principal exists.
\item The principal has the right to require information from A. This is the
case when a principal is either \texttt{responsible} for a component or 
\texttt{constructed} a component. 
\item The principal can sanction the component if it fails to give an
account. 
\end{itemize}

The last point is the most difficult to translate into the technical domain
because machines cannot be sanctioned, i.e., they feel neither remorse nor pain.
Thus, we translate this point to mean that a principal either
\texttt{constructed} a component and can thus  also change the component's
future behavior, access all logs and data, or that a principal is
\texttt{responsible} for a machine and thus knows about what the machine does
and can take some actions.  Finally, we formulate the generic definition given
by Hall et al. \cite{hall2017psy_acc} in psychology: A principal is
\texttt{hall\_accountable} for some component if the component's actions might
be evaluated by some principal. This can be expressed by using the informed
relation defined above. 
Note that while \texttt{hall\_accountable} and \texttt{lindberg\_accountable}
focus on the component, \texttt{raci\_accountable} focuses on the event. This
difference stems from the different perspective of the underlying theories,
i.e., social theories focus on the individual and thus on the component, and
RACI focuses on events in an organization.  Another noteworthy point is that
\texttt{raci-} and \texttt{lindberg\_accountable} consider only one principal
for a component, whereas \texttt{hall\_accountable} is a set.

\subsection{State Space}
\label{sec:state}

A cyber-physical system (\texttt{CPS}) is a collection of components and may
have a log file (multiple log files should be aggregated into a single logical
account). A CPS will naturally be able to perform actions that can affect its
surroundings. We decided not to model these actions explicitly because obtaining
such a list a~priori is typically impossible, having a list of potential actions
is not immediately useful, and all actions taken by the CPS should be reflected
in the log file.  Furthermore, each CPS has some associated principals and all
of its components require some configuration.

\begin{footnotesize}
\begin{schema}{CPS}
    system: \power Component\\
    logs: \power Observation\\
    principals : \power Principal\\
    setups: \power ComponentConfiguration\\
\where
system = ran(dom(logs))\\
system \neq \emptyset \land principals \neq \emptyset \land 
setups \neq \emptyset\\
ran(setups) \subseteq principals \land system = dom(setups)\\
\end{schema}

\end{footnotesize}

Typically CPS are not considered on their own, i.e., they are considered as part
of a larger socio-technical system (STS) that encompasses the CPS and  its
surroundings.  In our model, an \texttt{STS} consists of the
\texttt{ego\_system}, which is the system whose point of view we consider, and
\texttt{foreign\_cps}, i.e., other CPSs in the universe of discourse. We model
them as a sequence, which implies that any \texttt{foreign\_cps} will have some
form of unique ID.  Furthermore, it includes \texttt{principals}, of which the
principals of the \texttt{ego\_system} and the \texttt{foreign\_cps} are a
subset, as well as other \texttt{Beings} that may affect the technical systems.

\begin{footnotesize}
\begin{schema}{STS}
    ego\_system:  CPS\\
    foreign\_cps : seq\ CPS\\
    principals: \power Principal\\
    beings: \power Being\\
\where
ego\_system \notin  ran(foreign\_cps)\\
ego\_system.principals \subseteq principals\\
\forall c :  ran(foreign\_cps) @  c.principals \subseteq principals\\
\end{schema}

\end{footnotesize}

An \texttt{AccountabilityMechanism} is constructed over an STS and extends it
with abilities to log events and reason about them. It contains the
\texttt{STS}, \texttt{Events} that can be observed by principals,
\texttt{Observations}, i.e., events observed by principals, \texttt{Accounts}
that contain a list of all observations  by principals, external sources or
CPSs, \texttt{knownAccounts}, i.e., a list of accounts a principal knows about
and can thus use for reasoning, \texttt{directCause}, which provides a list of
components that caused some event, and \texttt{correctionActions}, i.e., the
actions that can be taken by principals to prevent a CPS from doing something.
In addition, \texttt{missedByEgo} contains all events recorded somewhere but
which are not known to the \texttt{ego\_system}. This set should be as small as
possible because not sensing an event often leads to errors or unexpected
behaviors.

\begin{footnotesize}
\begin{schema}{AccountabilityMechanism}	
	STS\\
	events: \power Event\\
    observations : \power Observation\\
    accounts: \power Account\\
    knownAccounts: \power hasAccount\\
    directCause: \power caused\\
    correctionActions : \power correctionAction\\ 
	missedByEgo : \power Event\\
\where
ran(ego\_system.logs) \subseteq accounts\\
\forall c :  ran(foreign\_cps) @ ran(c.logs) \subseteq accounts \\
dom(hasAccount) \subseteq accounts\\
dom(knownAccounts) \subset accounts\\
missedByEgo = events \setminus dom(dom(ego\_system.logs))
\end{schema}

\end{footnotesize}

\section{Applying the Model}
\label{sec:example}
\subsection{Example}

Here, we  show how our model can help clarify accidents, such as the previously
mentioned deadly crash of an autonomous vehicle operated by Uber. At the time of
writing, the official investigation was still on going, and we do not intend in
any way to second guess its results.  The goal of this example is to show how
difficult it is to clearly state what ``accountable'' means and capture its
impact on the system design. To that end, we base our example on the Uber
accident and fill any gaps in real system data with assumptions. To keep the
example concise, we limit the number of components, principals, and other
elements. Note that a real system would be significantly more complex than this
example. Furthermore, we do not claim that our understanding of the state space
is complete.  The example illustrates how modeling allows us to be precise in
understanding accountability. As  notation, we use a pseudo Z syntax that uses
state spaces to show the potential values of the given sets and axioms.  These
values, particularly those for the accountability relation or even
\texttt{caused}, do not need to be provided; however, they can be computed
(e.g., \cite{accbench} for causality).  In this example, we focus on two
questions: (1)~``Is Uber or the safety driver accountable?'' and (2)~``Did it
matter if the LIDAR worked?''. The first question is a classic question when
discussing cooperative systems, and the second questions assumes that a working
LIDAR, unimpaired by light conditions, could have avoided the accident.  

\subsection{Realizing the Given Sets}
We can define the chassis of the car, the accompanying sensors, its AI control
software and manual override as \texttt{Components}.  The \texttt{Account} would
encompass system logs, any sensor data, such as the recently released video
\cite{ubercrash}, or even human observation. Note that the \texttt{Events}
depend on the exact implementation. We assume that processed sensor data, e.g.,
object detection, or human testimonies, such as ``there was a crash'' will be
such events.  Of course all  events must be converted to a common format, which
is a nontrivial task. Here, the \texttt{Principals} are Uber because they
programmed, built, and configured the car, and the safety driver, because, here,
we assume that she could always intervene and stop the car. In addition, other
companies, like Volvo, who built the chassis, could be the principal for some
component.  Finally, \texttt{Actions} encompass countermeasures such as
``breaking to a full stop,'' ``swerve around the pedestrian,'' or similar
maneuvers. \texttt{Beings} include the pedestrian killed in the accident.

\subsection{Implementation-specific Interfaces}
An \texttt{Observation} in this context is a log entry about a specific event
(e.g., ``detect object'' or ``make a right turn'') or less formal knowledge such
as a person seeing something and  telling about it.  Finding useful
\texttt{correctionActions} is tricky and depends on what went wrong. Thus this
set may only be filled a~posteriori. In this example, we are seeking actions
that would avoid killing the pedestrian, e.g., ``break to a full stop'' or
``swerve right.'' Of course, these high-level actions can be broken down into
lower levels.  \texttt{hasAccount} returns the logs or information that each
principal possesses. In this example, it is easiest to prove what the safety
driver did and much more difficult to know what Uber did or did not do and,
crucially, at what time they knew. For the safety driver, we have her testimony
and the video released by the police department.  Here, \texttt{caused} would be
implemented with a reasoning algorithm or can be implemented by manually parsing
the log files. Finally, the \texttt{ComponentConfiguration} should be done by
Uber unless they outsourced some of their setup and configuration tasks. The
following state space  captures this example.

\begin{footnotesize}
\begin{schema}{exampleSTS}
STS\\
VIDEO,LIDAR, AI,CHASSIS, MANUAL\_CTRL : Component\\
DRIVER, UBER,VOLVO: Principal\\
BLACKBOX,VIDEO\_FEED,\\
DRIVER\_TESTEMONY: Account\\
PEDESTRIAN : Being\\
\where
ego\_system.system =  \{VIDEO,LIDAR,USONIC,AI \}\\
ego\_system.logs = \\
\{(DETECT\_PEDESTRIAN,LIDAR) \mapsto AI\}\\
ego\_system.principals = \{DRIVER, UBER\}\\
principals =  ego\_system.principals \\
ego\_system.setups = \{
VIDEO \mapsto UBER, LIDAR \mapsto UBER, \\
USONIC \mapsto UBER, AI \mapsto UBER, CHASSIS \mapsto VOLVO\}\\
foreign\_cps = \langle  \rangle \land beings = \{PEDESTRIAN\}\\
\end{schema}

\end{footnotesize}

\subsection{Context-independent Axioms}
The axioms  \texttt{informed}, \texttt{responsible}, \texttt{constructed},
\texttt{raci\_accountable} and \texttt{lindberg\-\_accountable} can be
realized on top of the underspecified axioms. If implemented according to the
specification, they yield useful and informed results about the
accountability within the system. 

\subsection{Notions of Accountability}
The formal model highlights that
\texttt{hall\_accountable}, \texttt{raci\_accountable} and
\texttt{lindberg\-\_accountable} require different implementation-specific
interfaces. To illustrate this difference, we first construct a simple
\texttt{AccountabilityMechanism} state space that has just enough information to
tell us which components are \texttt{hall\_accountable}.

In the first state space, we can infer $\mathit{hall\_accountable =
\{CHASSIS,AI\}}$, because events relating to these entities are known to some
principal. This form of accountability simply requires the \texttt{informed}
relation, which, in an implementation, means that some form of logging is
present. Note that there is no requirement for the form or quality of such logs.
It is considered accountable, as long as it is possible for a component to be
evaluated by some principal. 

\begin{footnotesize}
\begin{schema}{exampleAMhall}
	AccountabilityMechanism\\
	exampleSTS\\
	DETECT\_PEDESTRIAN, HIT\_PEDESTRIAN : Event\\
	BREAK, SWERVE : Action\\
\where
events = \{DETECT\_PEDESTRIAN, HIT\_PEDESTRIAN\}\\
observations =\\
\{(HIT\_PEDESTRIAN,CHASSIS) \mapsto   DRIVER\_TESTEMONY,\\
(HIT\_PEDESTRIAN,CHASSIS) \mapsto VIDEO\_FEED,\\
(DETECT\_PEDESTRIAN,LIDAR) \mapsto BLACKBOX
\}\\
accounts = ran(ego\_system.logs) \cup \{VIDEO\_FEED\}\\	
knownAccounts = \{BLACKBOX \mapsto UBER,\\
 VIDEO\_FEED \mapsto UBER,\\
 DRIVER\_TESTEMONY \mapsto DRIVER \}\\
informed(CHASSIS) = \{UBER,DRIVER\}\\
informed(AI) = \{UBER\}
\end{schema}

\end{footnotesize}

The next state space shows that, to determine whether there is a
\texttt{lindberg\_accountable} principle for some component, we also require a
mechanism to fill the set \texttt{correctionAction}. This is typically performed
manually after an accident has occurred. For future systems, such mechanisms
could be part of any explanation framework for an AI
\cite{doshi2017accountability}. In this example, we assume that the driver could
have paid better attention and stopped the car in time, and that the AI could
have reacted sooner than it did. Here, we can infer
$\mathit{lindberg\_accountable(CHASSIS) = DRIVER}$ and
$\mathit{lindberg\_accountable(AI) = UBER}$.

\begin{footnotesize}
\begin{schema}{exampleAMlindberg}
exampleAMhall
\where
correctionAction =\\
\{ (DRIVER,MANUAL\_CTRL) \mapsto BREAK,\\
 (UBER,AI) \mapsto BREAK \} \\
 responsible(AI) = UBER\\
 responsible(CHASSIS) = DRIVER
 \end{schema}

\end{footnotesize}

The final example requires that an even more complex operation, i.e.,
\texttt{caused}, is provided by the system. While currently causal relationships
are determined by experts, research in the field of causality suggests that
automated reasoning is possible \cite{halpern2015}. Here we incorporate the fact
that  Uber supposedly used too few LIDAR sensors to detect the pedestrian
\cite{lidar}. To reflect this knowledge, we remove \texttt{DETECT\_PEDESTRIAN}
from the system logs. This leads to \texttt{missedByEgo} to hold that event. A
causal reasoning algorithm can then conclude that the lack of detection led to
the collision and identify the \texttt{AI} and the \texttt{CHASSIS} as causes.
Thus, $\mathit{raci\_accountable(HIT\_PEDESTRIAN) = UBER}$ would then point to
Uber as the accountable entity. 

\begin{footnotesize}
\begin{schema}{exampleAMraci}
exampleAMlindberg
\where
ego\_system.logs =\\
\{(HIT\_PEDESTRIAN,CHASSIS) \mapsto BLACKBOX\}\\
events = \{DETECT\_PEDESTRIAN, HIT\_PEDESTRIAN\}\\
missedByEgo = \{DETECT\_PEDESTRIAN\}\\
directCause = \{HIT\_PEDESTRIAN \mapsto \{CHASSIS, AI \} \}\\
caused(HIT\_PEDESTRIAN) = \{AI,CHASSIS\}
\end{schema}

\end{footnotesize}
\subsection{Discussion}

Looking at the above examples, we see that both Uber and the safety driver are
accountable for some components. The reason for this ambiguity is that in our
model, the safety driver could always avoid the accident by being alert,
observing the pedestrian and breaking in time. This tacitly ignores the fact
that to avoid \texttt{HIT\_PEDESTRIAN}, a principal would first need to detect
her. This points us directly to the crux of self-driving cars, i.e., asking
humans to monitor a system for hours on end without anything happening, in the
hope they will react correctly in the split seconds leading up to an accident is
infeasible.  Humans, especially people without proper training, will invariably
lose focus and their attention will wander. In our example,
\texttt{DETECT\_PEDESTRIAN}  is only observed by the LIDAR, which is therefore
the only component that had the option to execute corrective actions. This does
not change the attribution of accountability for \texttt{hall\_accountable}, but
immediately changes \texttt{lindberg\_accountable} to exclude the driver because
she could not have done anything. \texttt{raci\_accountable} still depends on
the exact implementation of \texttt{caused}, which should include the new
knowledge to show that the driver is not accountable. Generally, our model
facilitates the comparison of the state of the world and the system's knowledge
of it.  Of course, there are many other ways to model the exact circumstances of
the accident, as well as many more notions of accountability. However, our goal
is not to find the one correct model or definition of accountability. On the
contrary, we strongly assume that there can be no single right answer. However,
we believe that a precise and formal model to state any assumptions and
explicate the notion of accountability used by the developers of a system is
useful and necessary. This will allow regulatory bodies to publish their
requirements in unambiguous language and make it possible to verify whether a
system is compliant or not.

\section{Conclusions}
In this paper we illustrated the benefits of a formal model of accountability
for cyber-physical systems. Our  model allows us to precisely describe the
notion of accountability a system should fulfill.  Such a formal description
enables us to analyze and compare these notions, and facilitates the selection
of the correct one for a system, thereby avoiding expensive, but superfluous,
subsystems and logging facilities.  In an over-simplified case study, we used
this  model to demonstrate the difference between three widely used notions of
accountability and described their impact on the implementation for our case
study. One clear area of future improvement is the fact that there are many more
notions of accountability in use today, which we seek to catalog, formalize, and
compare.  Having a unified view of accountability will help us identify the most
suitable notion for specific use cases.  Another widely open problem is how to
reason about such data.  Relations like \texttt{caused} or explanations for AI
systems are simple to ask for, but very difficult to realize in real-world
systems.  Furthermore, using terms like \emph{accountability},
\emph{responsibility} or \emph{causality} creates  clear connections to the
legal domain. Thus, it remains an open problem to characterize the connection
between legal terms, their technical meaning, and their application.  Ideally,
following a formal model of accountability will provide guarantees of legal
compliance.

\bibliographystyle{IEEEtran}
\bibliography{bibliography}

\end{document}